\documentclass[12pt]{article}

\def\gtwid{\mathrel{\raise.3ex\hbox{$>$\kern-.75em\lower1ex\hbox{$\sim$}}}}
\def\ltwid{\mathrel{\raise.3ex\hbox{$<$\kern-.75em\lower1ex\hbox{$\sim$}}}}
\def\square{\kern1pt\vbox{\hrule height 1.2pt\hbox{\vrule width 1.2pt\hskip 3pt
   \vbox{\vskip 6pt}\hskip 3pt\vrule width 0.6pt}\hrule height 0.6pt}\kern1pt}

\begin{document}

\begin{titlepage}

\begin{flushright}
UFIFT-QG-14-02
\end{flushright}

\vskip 2cm

\begin{center}
{\bf Nonlocal Metric Realizations of MOND}
\end{center}

\vskip 2cm

\begin{center}
R. P. Woodard$^{\dagger}$
\end{center}

\begin{center}
\it{Department of Physics, University of Florida, Gainesville, FL 32611}
\end{center}

\vspace{1cm}

\begin{center}
ABSTRACT
\end{center}

I discuss relativistic extensions of MOND in which the metric couples 
normally to matter. I argue that MOND might be a residual effect from 
the vacuum polarization of infrared gravitons produced during primordial
inflation. If so, MOND corrections to the gravitational field equations
would be nonlocal. Nonocality also results when one constructs metric 
field equations which reproduce the Tully-Fisher relation, along with 
sufficient weak lensing. I give the full field equations for the 
simplest class of models, and I specialize these equations to the 
geometries relevant for cosmology. I conclude by sketching the direction
of future studies.

\begin{flushleft}
\end{flushleft}

\vskip 2cm

\begin{flushleft}
$^{\dagger}$ e-mail: woodard@phys.ufl.edu
\end{flushleft}

\end{titlepage}

\section{Introduction}\label{intro}

Milgrom's MOdified Newtonian Dynamics (MOND) 
\cite{Milgrom:1983ca,Milgrom:1983pn,Milgrom:1983zz} has been 
wonderfully successful in explaining galactic structure without the 
need for dark matter \cite{Sanders:2002pf}. I work in quantum gravity, 
not astronomy, but I cannot forbear to present a list that Bob Sanders
compiled of the regularities of rotationally supported systems which MOND 
explains and which would otherwise need to be accidents of galaxy 
formation \cite{Sanders:2008iy}:
\begin{itemize}
\item{The Baryonic Tully-Fisher Relation $v_{\infty}^4
= a_0 G M$ between the asymptotic rotational velocity $v_{\infty}$, the
total mass $M$ and the MOND acceleration $a_0$, which holds over {\it 
five decades} in mass \cite{McGaugh:2005qe};}
\item{Milgrom's Law, that the need for dark matter (or MOND) seems
always to occur when the Newtonian acceleration drops below
$a_0$, \cite{Kaplinghat:2001me};}
\item{Freeman's Law $G\Sigma < a_0$ for the surface density $\Sigma$ of
a rotationally supported system \cite{Freeman:1970mx}; and}
\item{Sancisi's Law that for every feature in the surface brightness there
is a corresponding feature in the rotation curve, and vice versa 
\cite{Sancisi:2003xt}.}
\end{itemize}
Bob has an equally impressive list of successes for pressure-supported systems 
\cite{Sanders:2008iy} but I will let him make that case. The bottom line 
for a fundamental theorist like me is that MOND works better than anyone 
would expect if dark matter were the actual determinant of galactic 
structure.

Just as Newtonian dynamics is the nonrelativistic limit of general relativity,
so MOND must be the nonrelativistic limit of some larger theory. We need that
larger theory if we are to understand whether or not it can accomplish all the
things that dark matter does for general relativity. The search for an 
extension of MOND began early with the development of an action for the 
MOND-corrected Newtonian potential \cite{Bekenstein:1984tv}. Although not 
relativistic, this model demonstrates that MOND conserves energy, 3-momentum
and angular momentum.

There are now two classes of fully relativistic models:
\begin{itemize}
\item{Those in which the extra, MOND force is carried by some field other than
the metric; and}
\item{Those in which the MOND force is carried by the metric.}
\end{itemize}
Examples of the former are Bekenstein's TeVeS \cite{Bekenstein:2004ne} and
Moffat's STVG \cite{Moffat:2005si}. TeVeS has been studied extensively by 
some of the world's top cosmologists \cite{Skordis:2005xk,Dodelson:2006zt} 
and, while not without problems \cite{Clifton:2011jh}, it agrees better with 
observation than critics of MOND thought possible before the first relativistic 
extension was available for study. STVG (now called MOG) has not received 
independent attention but interesting results have been claimed for it 
\cite{Moffat:2012wn,Moffat:2013sja,Moffat:2013uaa}. 

When the MOND force is carried by fields other than the metric it raises 
the philosophical question of whether or not these other fields are a form of 
dark matter in disguise. Another peculiar feature of this class of models is 
that gravitons and matter particles propagate along the geodesics of different
metrics. This leads to an easily observable time lag between the arrival of 
the gravity wave pulse from some cosmic event (such as a supernova) and the 
corresponding optical and neutrino signal 
\cite{Kahya:2007zy,Desai:2008vj,Kahya:2008pp,Kahya:2010dk}. A single coincident 
detection of gravitational and optical (or neutrino) radiation from such an 
event would rule these models out.

If the MOND force is carried by the normal metric then the gravitational field
equations must take the form,
\begin{equation}
G_{\mu\nu} + \Delta G_{\mu\nu} = 8 \pi G T_{\mu\nu} \; , \label{ansatz}
\end{equation}
where $G_{\mu\nu} = R_{\mu\nu} - \frac12 g_{\mu\nu} R$ is the Einstein tensor,
$T_{\mu\nu}$ is the normal stress-energy tensor, and $\Delta G_{\mu\nu}$ 
represents the MOND correction. It has been difficult to build models of this
type because the Tully-Fisher relation precludes MOND corrections which derive 
from a local, invariant action. To see why, let us specialize to a static, 
spherically symmetric geometry in the weak field regime,
\begin{equation}
ds^2 = -\Bigl[1 + b(r)\Bigr] dt^2 + \Bigl[1 + a(r)\Bigr] dr^2 + r^2 d\Omega^2
\; . \label{statspher}
\end{equation}
Suppose the mass density $\rho(r)$ is always within the MOND regime of 
$\frac{G M(r)}{r^2} < a_0$, where $M(r)$ is the mass enclosed within radius $r$,
\begin{equation}
M(r) \equiv 4\pi \!\! \int_0^r \!\! dr' {r'}^2 \rho(r') \; .
\end{equation}
These conditions would pertain throughout a low surface brightness galaxy. 
For particles moving with $v(r)$ around circular orbits of radius $r$, the 
geodesic equation implies $v^2 = r b'/2$. Hence the Tully-Fisher relation
becomes,
\begin{equation}
\Bigl( \frac12 r b'(r)\Bigr)^2 = a_0 G M(r) \; . \label{Tully1}
\end{equation}
We can compare with the $\mu = 0 = \nu$ component of (\ref{ansatz}) by 
differentiating with respect to $r$ and then dividing by $\frac12 a_0 r^2$
\cite{Deffayet:2011sk},
\begin{equation}
\frac1{2 a_0 r^2} \frac{d}{dr} \Bigl( r b'(r)\Bigr)^2 = 8 \pi G \rho(r) \; .
\label{Tully2}
\end{equation}
The factors of $r$ are simple to understand in terms the spherical coordinate 
measure but a striking property of the left hand --- gravitational --- side 
of equation (\ref{Tully2}) is the presence of {\it three derivatives}. 
Local curvature invariants, and invariant derivatives of local curvatures,
contain even, nonzero, numbers of derivatives --- 2, 4, 6 and so on --- so
variations of them cannot produce the three derivatives which MOND 
phenomenology clearly requires.

In addition to precluding a local action, expression (\ref{Tully2}) is
peculiar because the left hand side is {\it quadratic} in the weak field 
$b(r)$ \cite{Soussa:2003sc}. This means that the MOND correction 
$\Delta G_{\mu\nu}$ must cancel the linear term from the Einstein tensor 
$G_{\mu\nu}$ --- at least for the $\mu = 0 = \nu$ component --- and then 
substitute the quadratic term we see in (\ref{Tully2}). Abandoning locality
allows us to accomplish both things. The basic idea is that acting the
inverse of a second order differential operator on a curvature produces a
weak field term with no derivatives, so the desired MOND correction to the
gravitational Lagrangian can take the form,
\begin{equation}
\Delta \mathcal{L} = -\Biggl[ \partial \times \frac1{\partial^2} \Bigl(
{\rm Curvature} \Bigr) \Biggr]^2 + \frac1{a_0} \Biggl[ \partial \times
\frac1{\partial^2} \Bigl( {\rm Curvature} \Bigr) \Biggr]^3 + \dots
\end{equation}
The first attempt to construct such an action was able to recover the 
Tully-Fisher relation but failed to produce the MOND enhancement of weak 
lensing which the data demand if dark matter is absent \cite{Soussa:2003vv}. 
It was finally possible to accommodate both requirements by involving a 
more complicated curvature \cite{Deffayet:2011sk}. (See also 
\cite{Hehl:2008eu,Hehl:2009es,Blome:2010xn}.) The purpose of this article 
is to explain how models of this type might arise from fundamental theory, 
to give the field equations for a general metric, and to specialize these 
equations to cosmology.

Section \ref{vacpol} discusses how significant nonlocal corrections to the
effective field equations might arise from the vacuum polarization of 
infrared gravitons created during an extended phase of primordial inflation. 
Section \ref{blocks} describes the nonlocal structures from which a 
phenomenologically viable model can be constructed. Section \ref{implement}
gives the full field equations for a general metric, and their specialization
to cosmology. I also discuss the possibility of making the MOND acceleration
$a_0$ dynamical so that it changes with the cosmological expansion rate. My
conclusions comprise section \ref{concl}.

\section{Gravitational Vacuum Polarization}\label{vacpol}

Abandoning locality is a profound step that requires justification. I 
believe that justification can be found in the phenomenon of gravitational
vacuum polarization. To understand this I begin by reviewing electromagnetic 
polarization, both in a medium and in vacuum. I then explain why 
gravitational vacuum polarization is so small in flat space background, 
and why there should have been a much larger effect during primordial
inflation. I believe that MOND might be a residual consequence of this.

Even undergraduate physics students are familiar with the phenomenon of 
polarization in a medium. The medium contains a vast number of bound charges. 
When an electric field is applied, the positive charges tend to move with the 
field and the negative changes move opposite. That charge separation 
polarizes the medium and tends to reduce the electric field strength.

One of the amazing predictions of quantum field theory is that virtual 
particles are continually emerging from the vacuum, existing for a brief
period, and then disappearing. How long these virtual particles can exist
is controlled by the energy-time uncertainty principle, which gives the 
minimum time $\Delta t$ needed to resolve and energy difference $\Delta E$,
\begin{equation}
\Delta t \Delta E \gtwid \hbar \; . \label{ET}
\end{equation}
(I will henceforth work in units where Planck's constant $\hbar$ and the 
speed of light $c$ are both unity.) If one imagines the emergence of a pair 
of positive and negatively charged particles of mass $m$ and momentum $\pm 
\vec{k}$ then the energy went from zero to $E = 2 [m^2 + k^2]^{\frac12}$. 
To {\it not} resolve a violation of energy conservation, the energy-time
uncertainty principle requires the pair to disappear after a time $\Delta t$ 
given by,
\begin{equation}
\Delta t \sim \frac1{\sqrt{m^2 + k^2}} \; . \label{flatDt}
\end{equation}
The rest is an exercise is classical (that is, non-quantum) physics. If we
ignore the change in the particles' momentum then their positions obey,
\begin{equation}
\frac{d^2}{dt^2} \Bigl( \sqrt{m^2 + k^2} \Delta \vec{x}_{\pm}\Bigr) =
\pm e \vec{E} \qquad \Longrightarrow \qquad \Delta \vec{x}_{\pm}(\Delta t)
= \frac{\pm e \vec{E}}{2 [m^2 + k^2]^{\frac32}} \; .
\end{equation}
Hence the polarization induced by wave vector $\vec{k}$ is,
\begin{equation}
\vec{p} = +e \Delta \vec{x}_+(\Delta t) - e \Delta \vec{x}_-(\Delta t) =
\frac{e^2 \vec{E}}{[m^2 + k^2]^{\frac32}} \; .
\end{equation}
The full vacuum polarization density comes from integrating $d^3k/(2\pi)^3$.

The simple analysis I have just sketched gives pretty nearly the prediction 
from one loop quantum electrodynamics, which is in quantitative agreement
with experiment. It allows us to understand two features of vacuum 
polarization which would be otherwise obscure:
\begin{itemize}
\item{That the largest effect derives from the lightest charged particles 
because they have the longest persistence times $\Delta t$ and therefore
induce the greatest polarization; and}
\item{That the electrodynamic interaction becomes stronger at short distances
because the longest wave length (hence smallest $k$) virtual particles could 
induce more polarization than is allowed by the travel time between two very
close sources.}
\end{itemize}
We can understand one more thing of great significance for the point I wish 
to make in this section: {\it vacuum polarization makes nonlocal corrections 
to the field equations and it would be a macroscopic phenomenon if there were 
massless charged particles.} In flat space background the quantum-corrected
Maxwell equations take the form \cite{Prokopec:2003bx}
\begin{equation}
\partial_{\nu} F^{\nu\mu}(x) - \partial_{\nu} \!\! \int \!\! d^4x' \chi_e(x;x')
F^{\nu\mu}(x') = J^{\mu}(x) \; , \label{Qmax}
\end{equation}
where $\chi_e(x;x')$ is the vacuum susceptibility. The one loop result for
massless, scalar quantum electrodynamics is \cite{Prokopec:2002uw},
\begin{equation}
\chi_e(x;x') = \frac{ \alpha \partial^4}{96 \pi^2} \Biggl\{ \theta\Bigl(
\Delta t \!-\! \Delta x\Bigr) \Biggl[ \ln\Bigl[\mu^2 \Bigl(\Delta t^2 \!-\!
\Delta x^2)\Bigr] \!-\! 1\Biggr] \Biggr\} + O(\alpha^2) \; ,
\end{equation}
where $\alpha$ is the fine structure constant, $\Delta t \equiv t - t'$, 
$\Delta x \equiv \Vert \vec{x} - \vec{x}'\Vert$ and $\mu$ is the renormalization 
scale. Note the causality which is manifest in the factor of $\theta(\Delta t
- \Delta x)$. By solving (\ref{Qmax}) for a point charge it is easy to show that 
one loop corrections screen the charge on distances $r > 1/\mu$ by a factor
which grows like $\ln(2 \mu r)$ \cite{Degueldre:2013hba}. In fact the effect
becomes so large that perturbation theory breaks if the interaction remains on
for too long.

The source of gravitation is stress-energy, and unlike electromagnetism it 
has only one sign. This means that the gravitational analogue of polarization
strengthens gravity, rather than weakening it. The application of a 
gravitational field to a classical medium attracts the material of the medium 
towards the gravitating source, which increases its gravitational field. 
Although gravitons are massless, and quantum gravitational vacuum polarization 
makes nonlocal corrections to the effective field equations, macroscopic effects 
are suppressed in flat space background because the ``charge'' of a graviton with 
wave vector $\vec{k}$ is $k$, which goes to zero for long wave length gravitons. 

The situation can be very different in an expanding universe. The associated 
geometry is characterized by a time dependent scale factor $a(t)$,
\begin{equation}
ds^2 \equiv g_{\mu\nu} dx^{\mu} dx^{\nu} = -dt^2 + a^2(t) d\vec{x} \!\cdot\!
d\vec{x} \; . \label{FRW}
\end{equation}
Two derivatives of $a(t)$ characterize the expansion,
\begin{equation}
H(t) \equiv \frac{\dot{a}}{a} \qquad , \qquad \epsilon(t) \equiv -
\frac{\dot{H}}{H^2} \; .
\end{equation}
The Hubble parameter $H(t)$ gives the rate of cosmological expansion, and the 
first slow roll parameter $\epsilon(t)$ provides a dimensionless measure of 
how fast that rate is changing. The crucial boundary for this discussion is
$\epsilon = 1$. For larger values of $\epsilon$ the universe is said to be
decelerating because $\ddot{a} < 0$, whereas $\epsilon < 1$ corresponds to 
accelerated expansion ($\ddot{a} > 0$) or inflation. We know inflation can
happen because it is going on right now, as witness the current values of
$H$ and $\epsilon$ \cite{Ade:2013zuv},
\begin{equation}
H_0 \approx 2.2 \times 10^{-18}~{\rm Hz} \qquad , \qquad \epsilon_0 \approx 
.47 \; . \label{current}
\end{equation}
However, the period of relevance to this discussion is the epoch of primordial 
inflation, which is thought to have occurred at about $10^{-32}$ seconds after 
the Big Bang. If the BICEP detection of primordial $B$-mode polarization is 
correct then the values of $H$ and $\epsilon$ at that time were
\cite{Ade:2014xna},
\begin{equation}
H_i \approx 1.8 \times 10^{38}~{\rm Hz} \qquad , \qquad 
\epsilon_i \approx 0.013 \; . \label{primordial}
\end{equation}
These values tell us two crucial things about primordial inflation:
\begin{itemize}
\item{The quantum gravitational loop counting parameter of $(\hbar/c^5) 
G H_i^2 \approx 7.7 \times 10^{-9}$ is small enough that perturbation theory
is valid, but not so small that quantum gravitational effects are negligible;
and}
\item{The slow roll parameter is close to the de Sitter value of $\epsilon = 
0$, at which point the Hubble parameter becomes constant and the scale factor
is $a(t) \approx a_i e^{H_i t}$.}
\end{itemize}

Cosmological expansion can strengthen quantum effects because it causes the
virtual particles which drive them to persist longer. This is easy to see 
from the geometry (\ref{FRW}). Because spatial translation invariance is
unbroken, particles still have conserved wave numbers $\vec{k}$. However,
because the physical distance is the coordinate distance scaled by $a(t)$,
the physical energy of a particle with mass $m$ and wave number $k = 
2\pi/\lambda$ becomes time dependent,
\begin{equation}
E(t,\vec{k}) = \sqrt{m^2 + \frac{k^2}{a^2(t)} } \; .
\end{equation}
Hence the relation for the persistence time $\Delta t$ of a virtual pair 
which emerges at time $t$ changes from (\ref{flatDt}) to,
\begin{equation}
\int_{t}^{t+\Delta t} \!\!\!\! dt' E(t',\vec{k}) \sim 1 \; . \label{expDt}
\end{equation}
Massless particles persist the longest, just as they do in flat space. 
However, for inflation it is the lower limit of (\ref{expDt}) which 
dominates, so that even taking $\Delta t$ to infinity does not cause the
integral to grow past a certain point. One can see this from the de Sitter 
limit,
\begin{equation}
\int_{t}^{t+\Delta t} \!\!\!\! dt' \frac{k}{a(t')} = \frac{k}{H_i a(t)}
\Bigl[ 1 - e^{-H_i \Delta t}\Bigr] \; . \label{inflDt}
\end{equation}
A particle with $k < H(t) a(t)$ is said to be super-horizon, and we have
just shown that {\it any massless virtual particle which emerges from the 
vacuum with a super-horizon wave number during inflation will persist forever.}

It turns out that almost all massless particles possess a symmetry known as 
{\it conformal invariance} which suppresses the rate at which they emerge from
the vacuum. This keeps the density of virtual particles small, even though any 
that do emerge can persist forever. One can see the problem by specializing 
the Lagrangian of a massless, conformally coupled scalar $\psi(t,\vec{x})$ to 
the cosmological geometry (\ref{FRW}),
\begin{equation}
\mathcal{L} = -\frac12 \partial_{\mu} \psi \partial_{\nu} \psi g^{\mu\nu} 
\sqrt{-g} -\frac{R}{12} \psi^2 \sqrt{-g} \longrightarrow \frac{a^3}{2} \Bigl[
\dot{\psi}^2 - \frac{\partial_i \psi \partial_i \psi}{a^2} - (\dot{H} + 2 H^2) 
\psi^2\Bigr] \; .
\end{equation}
The equation for a canonically normalized, spatial plane wave of the form 
$\psi(t,\vec{x}) = v(t,k) e^{i\vec{k} \cdot \vec{x}}$ can be solved for a 
general scale factor $a(t)$,
\begin{equation}
\ddot{v} + 3 H \dot{v} + \Bigl[ \frac{k^2}{a^2} + \dot{H} + 2 H^2\Bigr]
v = 0 \; \Longrightarrow \; v(t,k) = \frac1{a(t) \sqrt{2k}}
\exp\Bigl[ -ik \!\! \int_{t_i}^t \!\! \frac{dt'}{a(t')} \Bigr] \; . 
\label{vmode}
\end{equation}
The factor of $1/a(t)$ in (\ref{vmode}) suppresses the emergence rate,
even though destructive interference from the phase dies off, just as the 
energy-time uncertainty principle (\ref{inflDt}) predicts. The stress-energy 
contributed by this field is,
\begin{equation}
T_{\mu\nu} = \Bigl[\delta^{\rho}_{\mu} \delta^{\sigma}_{\nu} \!-\! \frac12 
g_{\mu\nu} g^{\rho\sigma} \Bigr] \partial_{\rho} \psi \partial_{\sigma} \psi
+ \frac16 \Bigl[ R_{\mu\nu} \!-\! \frac12 g_{\mu\nu} R \!+\! g_{\mu\nu} 
\square \!-\! D_{\mu} D_{\nu} \Bigr] \psi^2 \; ,
\end{equation}
where $D_{\mu}$ is the covariant derivative and $\square$ is the covariant 
d'Alembertian. We can get the 0-point energy of a single wave vector $\vec{k}$
by specializing $T_{00}$ to the cosmological geometry (\ref{FRW}) and multipling 
by a factor of $a^3(t)$,
\begin{equation}
\mathcal{E}(t,\vec{k}) = \frac{a^3}{2} \Bigl[ \vert \dot{v}\vert^2 + \Bigl( 
\frac{k^2}{a^2} \!+\! H^2\Bigr) \vert v\vert^2 + H \Bigl( v \dot{v}^* \!+\! 
\dot{v} v^*\Bigr) \Bigr] = \frac{k}{2 a(t)} \; . \label{confE}
\end{equation}
This is just the usual $\frac12 \hbar \omega$ term which is not strengthened 
but rather weakened by the cosmological expansion.

Only gravitons and massless, minimally coupled scalars are both massless and 
not conformally invariant so that they can engender significant quantum effects 
during inflation. Because they obey the same mode equation 
\cite{Lifshitz:1945du,Grishchuk:1974ny} it will suffice to specialize the 
scalar Lagrangian to the cosmological geometry (\ref{FRW}),
\begin{equation}
\mathcal{L} = -\frac12 \partial_{\mu} \phi \partial_{\nu} \phi g^{\mu\nu} 
\sqrt{-g} \longrightarrow \frac12 a^3 \Bigl[\dot{\phi}^2 - \frac1{a^2}
\partial_i \phi \partial_i \phi \Bigr] \; .
\end{equation} 
The equation for a canonically normalized, spatial plane wave of the form 
$\phi(t,\vec{x}) = u(t,k) e^{i\vec{k} \cdot \vec{x}}$ is simpler than that
of its conformally coupled cousin (\ref{vmode}) but more difficult to solve,
so I will specialize the solution to de Sitter,
\begin{equation}
\ddot{u} + 3 H \dot{u} + \frac{k^2}{a^2} \, u = 0 \; \Longrightarrow \; 
u(t,k) = \frac1{a(t) \sqrt{2k}} \Bigl[1 + \frac{i H_i a(t)}{k}\Bigr]
\exp\Bigl[ -ik \!\! \int_{t_i}^t \!\! \frac{dt'}{a(t')} \Bigr] \; . 
\label{umode}
\end{equation}
The minimally coupled mode function $u(t,k)$ has the same phase factor
as the conformal mode function (\ref{vmode}), and they both fall off
like $1/a(t)$ in the far sub-horizon regime of $k \gg H_i a(t)$. 
However, they disagree strongly in the super-horizon regime during which
$v(t,k)$ continues to fall off whereas $u(t,k)$ approaches a phase times
$H_i/\sqrt{2 k^3}$. One can see from the equation on the left of
(\ref{umode}) that $u(t,k)$ approaches a constant for any inflating
geometry. 

The 0-point energy in wave vector $\vec{k}$ is,
\begin{equation}
\mathcal{E}(t,\vec{k}) = \frac12 a^3 \Bigl[ \vert \dot{u}\vert^2 +
\frac{k^2}{a^2} \vert u\vert^2 \Bigr] = \frac{k}{a(t)} \Bigl[ \frac12
+ \Bigl( \frac{H_i a(t)}{2 k} \Bigr)^2 \Bigr] \; . \label{E/k}
\end{equation}
Because each wave vector is an independent harmonic oscillator with 
mass proportional to $a^3(t)$ and frequency $k/a(t)$ we can read off
the occupation number from expression (\ref{E/k}),
\begin{equation}
N(t,\vec{k}) = \Bigl[ \frac{H_i a(t)}{2 k} \Bigr]^2 \; . \label{N/k}
\end{equation}
As one might expect, this number is small in the sub-horizon regime.
It becomes of order one at the time $t_k$ of horizon crossing, $k =
H(t_k) a(t_k)$, and $N(t,\vec{k})$ grows explosively afterwards. This
is the basis of the power spectra of scalar \cite{Mukhanov:1981xt} and 
tensor \cite{Starobinsky:1979ty} perturbations which are predicted
by primordial inflation.

We can use the occupation number (\ref{N/k}) to compute the energy 
density of infrared ($k < H_i a(t)$) gravitons,
\begin{equation}
\rho_{\scriptscriptstyle {\rm IR}} = \int \!\! \frac{d^3k}{(2 \pi a)^3}
\theta\Bigl( H_i a \!-\! k\Bigr) \times 2 \times N(t,\vec{k}) \times 
\frac{k}{a} = \frac{H_i^4}{8\pi^2} \; .
\end{equation}
This is smaller than the energy density which caused inflation by a 
factor of $G H_i^2/3\pi \sim 10^{-9}$, but it is still a huge energy
density by today's standards, and one can easily see that the continual
particle production needed to maintain $\rho_{\scriptscriptstyle 
{\rm IR}}$ must limit the duration of inflation. To keep things finite, 
suppose that the spatial topology is that of a 3-torus, with radius 
$R(t) = a(t)/H_i$. If de Sitter expansion proceeds unimpeded then the 
total mass at time $t$ is,
\begin{equation}
M(t) \sim \rho_{\scriptscriptstyle {\rm IR}} R^3(t) \sim H_i a^3(t) \; .
\end{equation}
If this mass were in causal contact with itself then the universe 
would fall within its own Schwarzschild radius at time $t$ such that,
\begin{equation}
1 \sim \frac{G M(t)}{R(t)} = G H_i^2 a^2(t) \; . \label{estimate}
\end{equation}
That corresponds to only about 9 e-foldings, which is far less than 
the 50 to 60 e-foldings of inflation which must occur to solve the 
horizon problem.

My estimate (\ref{estimate}) was based on assuming that all the mass 
is instantaneously in contact with itself. Of course causality imposes 
a time lag between when a particle first emerges from the vacuum and 
when its gravity can affect the particles which have already emerged.
{\it That time lag is the only thing holding the universe up against
gravitational collapse.} It seems obvious that causality cannot extend 
the duration of inflation forever. A simple of measure of how fast 
things grow causally is the volume of the past light-cone back to the
beginning of time. For flat space this goes like $t^4$, but its growth 
during inflation is only like $\ln[a(t)/a_i] \sim H_i t$. So one should 
expect the back-reaction from inflationary graviton production to become 
effective when $\ln[a(t)/a_i] \sim 1/G H_i^2$. 

With Nick Tsamis I have proposed a model in which inflation begins due 
to a large, positive cosmological constant that is screened by the 
gradual build-up of self-gravitation between inflationary gravitons 
\cite{Tsamis:1996qq,Tsamis:2011ep}. Direct computations of quantum 
gravity on de Sitter background are difficult, and plagued by problems 
of interpretation \cite{Garriga:2007zk,Tsamis:2008zz}, but they do show 
secular corrections of $\ln[a(t)/a_i]$ \cite{Miao:2005am,Miao:2006gj} 
according to well-defined rules \cite{Prokopec:2007ak}. These factors
eventually grow so large that perturbation theory breaks down, at which
point some sort of nonperturbative scheme must be employed to follow
subsequent evolution. Developing such a technique is not hopeless 
because Starobinsky and Yokoyama were able to accomplish it for scalar
potential models \cite{Starobinsky:1994bd,Tsamis:2005hd}, and the
technique has been generalized to scalars which interact with photons
\cite{Prokopec:2007ak} and with fermions \cite{Miao:2006pn}. However,
the generalization to quantum gravity has not been achieved yet
\cite{Miao:2008sp,Kitamoto:2010et,Kitamoto:2011yx}. 

In the absence of a nonperturbative resummation technique there have 
been efforts to use perturbative results to make a plausible guess for 
the most cosmologically significant part of the effective field equations 
\cite{Tsamis:1997rk,Tsamis:2009ja,Tsamis:2010ph}. The point of relevance
to MOND is that these models are necessarily nonlocal 
\cite{Romania:2012av}, and that they inevitably change the force of 
gravity. I will close this section by commenting that this physical 
picture helps to justify two features phenomenological models of MOND 
which would be otherwise inexplicable:
\begin{itemize}
\item{There is a beginning of time, corresponding roughly to the end of
inflation when quantum gravitational corrections became nonperturbatively 
strong; and}
\item{Because the mechanism derives from {\it cosmological} gravitons, we
should expect the largest modifications of gravity on large scales, not
on small scales.}
\end{itemize}

\section{Building Blocks for Nonlocal Models}\label{blocks}

It turns out that MOND phenomenology can be implemented using only two 
nonlocal structures \cite{Deffayet:2011sk}:
\begin{itemize}
\item{The inverse (with retarded boundary conditions) of the scalar 
d'Alembertian \cite{Tsamis:1997rk},
\begin{equation}
\square \equiv \frac1{\sqrt{-g}} \, \partial_{\mu} \Bigl( \sqrt{-g}
g^{\mu\nu} \partial_{\nu} \Bigr) \; ; \; {\rm and}
\end{equation}}
\item{The normalized gradient of some nonlocal scalar $\chi[g]$, such
as the volume of the past lightcone \cite{Park:2009ai}, which grows in 
the timelike direction,
\begin{equation}
u^{\mu}[g] \equiv \frac{-g^{\mu\nu} \partial_{\nu} \chi}{\sqrt{
-g^{\alpha\beta} \partial_{\alpha} \chi \partial_{\beta} \chi}} \; .
\end{equation}}
\end{itemize}
The apparent sufficiency of these quantities may result from adopting 
too narrow a focus; one should remain open to the possibility of higher 
spin operators \cite{Ferreira:2013tqn} and more general forms of 
nonlocality. In this section I specialize these quantities and some 
relevant curvatures to the static, spherically symmetric geometry 
(\ref{statspher}) and to the homogeneous, isotropic and spatially flat 
geometry (\ref{FRW}) of cosmology. It turns out that this can be done 
without giving a precise specification for the scalar $\chi[g]$.

Because I do not necessarily make the weak field approximation it is
useful to define $A(r) \equiv 1 + a(r)$ and $B(r) \equiv 1 + b(r)$ for
the static, spherically symmetric geometry. The scalar d'Alembertian
and normalized, timelike 4-velocity are,
\begin{equation}
\square F(r) = \frac1{r^2 \sqrt{AB}} \frac{d}{dr} \Biggl[ r^2 
\sqrt{\frac{B(r)}{A(r)}} F'(r) \Biggr] \qquad , \qquad
u^{\mu} = \frac{\delta^{\mu}_0}{\sqrt{B(r)}} \; .
\end{equation}
The $00$ component and trace of the Ricci tensor are,
\begin{eqnarray}
R_{00} & = & \frac{B''}{2A} - \frac{B'}{4 A} \Bigl( \frac{A'}{A}
\!+\! \frac{B'}{B}\Bigr) + \frac{B'}{r A} \; , \\
R & = & -\frac{B''}{AB} + \frac{B'}{2AB} \Bigl( \frac{A'}{A} \!+\!
\frac{B'}{B} \Bigr) + \frac2{r A} \Bigl( \frac{A'}{A} \!-\! \frac{B'}{B}
\Bigr) + \frac{2 (A \!-\! 1)}{r^2 A} \; . \qquad
\end{eqnarray}
Assuming regularity at $r=0$ one can recover the potential $B(r)$ from a 
nonlocal scalar,
\begin{equation}
\frac1{\square} \Bigl( u^{\alpha} u^{\beta} R_{\alpha\beta} \Bigr) 
= \frac12 \ln\Bigl[ \frac{B(r)}{B(0)}\Bigr] \; . \label{B(r)}
\end{equation}

For the cosmological geometry (\ref{FRW}) the scalar d'Alembertian and
the timelike 4-velocity take the forms,
\begin{equation}
\square F(t) = -\frac1{a^3(t)} \frac{d}{dt} \Bigl[ a^3(t) \dot{F}(t)\Bigr]
\qquad , \qquad u^{\mu} = \delta^{\mu}_0 \; .
\end{equation}
The $00$ component and the trace of the Ricci tensor are,
\begin{eqnarray}
R_{00} & = & -3 (\dot{H} \!+\! H^2) \; , \\
R & = & 6\dot{H} + 12 H^2 \; .
\end{eqnarray}
One can very largely reconstruct the scale factor from another nonlocal
scalar,
\begin{eqnarray}
\frac1{\square} \Bigl( R + u^{\alpha} u^{\beta} R_{\alpha\beta}\Bigr) 
& = & -3 \!\! \int_{t_i}^t \!\! \frac{dt'}{a^3(t')} \!\! \int_{t_i}^{t'}
\!\! dt'' \frac{d}{dt''} \Bigl( H(t'') a^3(t'') \Bigr) \; , \\
& = & -3 \ln\Bigl[ \frac{a(t)}{a_i}\Bigr] + 3 \!\! \int_{t_i}^t \!\! dt'
\frac{H_i a_i^3}{a^3(t')} \; .
\end{eqnarray}
Note the implementation of retarded boundary conditions.

\section{MOND Field Equations and Cosmology}\label{implement}

The static, spherically symmetric geometry (\ref{statspher}) has two 
gravitational potentials, $B(r)$ and $A(r)$. Requiring that the 
ultra-weak field limit of the $B(r)$ equation should reproduce the 
Tully-Fisher relation, along with no change in the $A(r)$ equation, 
restricts invariant, metric-based gravitational Lagrangians to take the 
form of general relativity plus \cite{Deffayet:2011sk},
\begin{equation}
\Delta \mathcal{L} = \frac{a_0^2}{16 \pi G} f\Bigl( \frac{Y[g]}{a_0^2}
\Bigr) \sqrt{-g} \; , \; Y[g] \equiv g^{\mu\nu} \partial_{\mu} 
\frac{2}{\square} \Bigl(u^{\alpha} u^{\beta} R_{\alpha\beta}\Bigr)
\partial_{\nu} \frac2{\square} \Bigl( u^{\rho} u^{\sigma} 
R_{\rho\sigma} \Bigr) \; . \label{general}
\end{equation} 
MOND phenomenology requires $f(Z) = \frac12 Z - \frac16 Z^{\frac32} + 
O(Z^2)$, with major suppression for large, positive $Z$. It does not fix
the behavior of $f(Z)$ for negative $Z$. The purpose of this section is
to give the general field equations for this class of models and then 
specialize them to the cosmological geometry (\ref{FRW}). For simplicity
I will assume that the nonlocal scalar $\chi[g]$ is,
\begin{equation}
\chi[g] \equiv -\frac1{\square} 1 \; . \label{chidef}
\end{equation}

An easy way of getting the field equations is through the auxiliary 
scalar formalism that Nojiri and Odintsov \cite{Nojiri:2007uq}
introduced to localize a nonlocal model of dark energy \cite{Deser:2007jk}.
The general MOND model (\ref{general}) requires four scalars: $\chi[g] = 
-\frac1{\square} 1$ and $\phi[g] = \frac2{\square} (u^{\alpha} u^{\beta} 
R_{\alpha\beta})$, along with Lagrange multiplier fields $\psi$ and $\xi$ 
to enforce these conditions. The localized form is \cite{Deffayet:2014XX},
\begin{eqnarray}
\lefteqn{\Delta \mathcal{L} = \frac1{16\pi G} \Biggl\{ a_0^2 f\Bigl(
\frac{g^{\mu\nu} \partial_{\mu} \phi \partial_{\nu} \phi}{a_0^2}
\Bigr) } \nonumber \\
& & \hspace{2cm} - \Bigl[ \partial_{\mu} \xi \partial_{\nu} \phi 
g^{\mu\nu} \!+\! 2 \xi R_{\mu\nu} u^{\mu} u^{\nu} \Bigr] - \Bigl[ 
\partial_{\mu} \psi \partial_{\nu} \chi g^{\mu\nu} \!-\! \psi\Bigr] 
\Biggr\} \sqrt{-g} \; . \qquad \label{local}
\end{eqnarray}
The MOND correction $\Delta G_{\mu\nu}$ to the Einstein tensor is 
\cite{Deffayet:2014XX},
\begin{eqnarray}
\lefteqn{ \frac{16 \pi G}{\sqrt{-g}} \frac{\delta \Delta S}{\delta
g^{\mu\nu}} = \frac12 g_{\mu\nu} \Bigl[ -a_0^2 f + g^{\rho\sigma}
\Bigl( \partial_{\rho} \xi \partial_{\sigma} \phi \!+\! \partial_{\rho}
\psi \partial_{\sigma} \chi\Bigr) + 2 \xi u^{\rho} u^{\sigma} 
R_{\rho\sigma} - \psi\Bigr] } \nonumber \\
& & \hspace{0cm} + \partial_{\mu} \phi \partial_{\nu} \phi f' -
\partial_{(\mu} \xi \partial_{\nu)} \phi - \partial_{(\mu} \psi 
\partial_{\nu)} \chi - 2 \xi \Bigl[ 2 u_{(\mu} u^{\alpha} R_{\nu ) \alpha}
\!+\! u_{\mu} u_{\nu} u^{\alpha} u^{\beta} R_{\alpha\beta} \Bigr] 
\nonumber \\
& & \hspace{2.5cm} - \Bigl[ \square (\xi u_{\mu} u_{\nu}) +
g_{\mu\nu} D_{\alpha} D_{\beta} (\xi u^{\alpha} u^{\beta}) - 
2 D_{\alpha} D_{(\mu} (\xi u_{\nu)} u^{\alpha} ) \Bigr] \; . \qquad
\label{DGmn}
\end{eqnarray}
Unlike Nojiri and Odintsov, we do not take the various scalars to have
independent initial value data (which would result in two linear 
combinations of them being ghosts \cite{Deser:2013uya,Woodard:2014iga})
but instead define each scalar using $\frac1{\square}$ with retarded 
boundary conditions \cite{Deffayet:2014XX},
\begin{eqnarray}
\phi[g] = \frac{2}{\square} \Bigl(u^{\alpha} u^{\beta} R_{\alpha\beta}
\Bigr) & , & \xi[g] = \frac{2}{\square} \Biggl(D_{\mu} \Biggl[ D^{\mu} 
\phi f'\Bigl( \frac{g^{\rho\sigma} \partial_{\rho} \phi 
\partial_{\sigma} \phi}{a_0^2} \Bigr) \Biggr] \Biggr) \; , 
\label{phixi} \\
\chi[g] = -\frac1{\square} \Bigl(1\Bigr) & , & \psi[g] = 
\frac{4}{\square} \Biggl(D_{\mu} \Biggl[ \frac{ \xi g^{\mu\rho}_{\perp} 
u^{\sigma} R_{\rho\sigma}}{\sqrt{-g^{\alpha\beta} \partial_{\alpha} 
\chi \partial_{\beta} \chi}} \Biggr] \Biggr) \; , \qquad \label{chipsi}
\end{eqnarray}
where $g^{\mu\nu}_{\perp} \equiv g^{\mu\nu} + u^{\mu} u^{\nu}$ is the
projected metric.

The scalars take simple forms in the cosmological geometry (\ref{FRW})
\cite{Deffayet:2014XX},
\begin{eqnarray}
\phi(t) = 6 \!\! \int_{t_i}^t \!\! \frac{dt'}{a^3(t')} \!\!
\int_{t_i}^{t'} \!\! dt'' a^3 \Bigl[\dot{H} + H^2\Bigr] & , & 
\xi(t) = 2 \!\! \int_{t_i}^t \!\! dt' \dot{\phi}(t') 
f'\Bigl(-\frac{\dot{\phi}^2(t')}{a_0^2} \Bigr) \; , \label{FLRWphixi} \\
\chi(t) = \!\! \int_{t_i}^t \!\! \frac{dt'}{a^3(t')} \!\!
\int_{t_i}^{t'} \!\! dt'' a^3(t'') & , & \psi(t) = 0 \; . 
\label{FLRWchipsi}
\end{eqnarray}
Homogeneity and isotropy imply that any second rank tensor 
such as $\Delta G_{\mu\nu}$ has only two distinct components when 
specialized to cosmology (\ref{FRW}) \cite{Deffayet:2014XX},
\begin{eqnarray}
\Delta G_{00}(t) & = & \frac{a_0^2}2 f\Bigl( \frac{-\dot{\phi}^2}{a_0^2} 
\Bigr) + 3 H \dot{\xi} + 6 H^2 \xi \; , \label{00eqn} \\
\Delta G_{ij}(t) & = & -\Biggl[\frac{a_0^2}2 f\Bigl(
\frac{-\dot{\phi}^2}{a_0^2}\Bigr) + \ddot{\xi} + (\frac{\dot{\phi}}2 
\!+\! 4 H) \dot{\xi} + (4\dot{H} \!+\! 6 H^2) \xi\Biggr] g_{ij} \; . 
\qquad \label{ijeqn} 
\end{eqnarray}

As one can see from expressions (\ref{FLRWphixi}) and (\ref{00eqn}-\ref{ijeqn}),
the MOND function $f(Z)$ is evaluated at a negative definite argument $Z = -
\dot{\phi}^2(t)/a_0^2$ for cosmology. MOND phenomenology only fixes the 
behavior of $f(Z)$ for positive $Z$ \cite{Deffayet:2011sk}. We are therefore
free to adjust the negative $Z$ branch so that the MOND additions to the 
Einstein tensor (\ref{00eqn}-\ref{ijeqn}) make up for the absence of dark 
matter in determining the expansion history $a(t)$. This is accomplished 
by regarding $a(t)$ as known, with the matter density $\rho$ also known in
terms of $a(t)$, and then treating the modified Friedmann equation 
$3 H^2 + \Delta G_{00} = 8 \pi G \rho$ as an integro-differential equation for 
$f(Z)$ which can be solved numerically \cite{Deffayet:2014XX}. A very similar 
exercise was worked out in detail \cite{Deffayet:2009ca} to determine the 
negative branch of the free function of nonlocal cosmology \cite{Deser:2007jk} 
so as to make this model reproduce the $\Lambda$CDM expansion history without 
a cosmological constant. Because the nonlocal MOND equations (\ref{general})
are fixed once the function $f(Z)$ has been defined, one can subject the model 
to meaningful tests by studying its response to perturbations around the 
cosmological geometry. Dodelson and Park have recently done this for nonlocal 
cosmology \cite{Park:2012cp,Dodelson:2013sma}, and one can directly apply
their treatment of the operator $1/\square$.

Although there seems to be no obstacle to reconstructing $f(Z)$ to make 
nonlocal MOND support the known expansion history without dark matter, a 
potential problem is the enormous range of $Z = -\dot{\phi}^2(t)/a_0^2$ that 
would be involved. This becomes apparent from the time derivative of 
(\ref{FLRWphixi}),
\begin{eqnarray}
\dot{\phi}(t) & = & -\frac{6}{a^3(t)} \!\! \int_{t_i}^t \!\! dt' a^3(t') 
H^2(t') (\epsilon \!-\! 1) \; , \\
& = & -6 H \Bigl( \frac{\epsilon \!-\! 1}{3 \!-\! \epsilon}\Bigr) +  
6 H_i \Bigl( \frac{\epsilon_i \!-\! 1}{3 \!-\! \epsilon_i}\Bigr) \Bigl(
\frac{a_i}{a}\Bigr)^3 + \frac{6}{a^3} \!\! \int_{t_i}^t \!\! dt' a^3 H 
\frac{d}{dt'} \Bigl( \frac{ \epsilon \!-\! 1}{3 \!-\! \epsilon} \Bigr) \; . 
\qquad \label{smalleps} 
\end{eqnarray}
Because the slow roll parameter $\epsilon(t) \equiv -\dot{H}/H^2$ has only
small temporal variation, the integral in expression (\ref{smalleps}) is
negligible and the first term dominates. It has been noted from
the earliest papers \cite{Milgrom:1983pn} that the MOND acceleration $a_0$ is
close to the current value of the Hubble parameter $H_0$ times the speed of 
light (which is one in my units). Hence we see that cosmological evolution 
results in staggering changes in $Z(t) \equiv -\dot{\phi}^2(t)/a_0^2 \sim 
-H^2(t)/H_0^2$. For example, it would be about $Z \sim -10^{32}$ during 
nucleosynthesis, and about $Z \sim -10^{10}$ at the time of recombination. 

There may be no problem with such dramatic variation in $Z(t)$, but it is 
surely worth considering the alternative of replacing the MOND acceleration 
$a_0$ by some dynamical quantity $\alpha[g]$ which changes with the expansion 
of spacetime. Although this step would make major changes in cosmology, there 
is no strong evidence for or against it from observable cosmic structures
\cite{Milgrom:2008rv,Milgrom:2008cs,Bekenstein:2008pc}. One of the simplest 
choices for $\alpha[g]$ is proportional to the local expansion derived from 
the divergence of the timelike 4-velocity $u^{\mu}[g]$ \cite{Deffayet:2014XX},
\begin{equation}
a_0 \longrightarrow \alpha[g] \equiv \frac{D_{\mu} u^{\mu}[g]}{6\pi} \; . 
\label{alpha}
\end{equation} 

The replacement (\ref{alpha}) engenders only three changes in the MOND
correction $\Delta G_{\mu\nu}[g]$ (\ref{DGmn}-\ref{chipsi}) for a general 
metric. Of course the factors of $a_0$ in (\ref{DGmn}) and (\ref{phixi})  
must be replaced by $\alpha[g]$. The second difference is that $\Delta 
G_{\mu\nu}$ acquires an extra contribution from the metric dependence of 
$\alpha[g]$ \cite{Deffayet:2014XX},
\begin{eqnarray}
\lefteqn{\Bigl(\Delta G_{\mu\nu} \Bigr)_{\rm new} = \Bigl( \Delta G_{\mu\nu} 
\Bigr)_{\rm old} + g_{\mu\nu} \Bigl[\alpha^2 f - g^{\rho\sigma} \partial_{\rho}
\phi \partial_{\sigma} \phi f'\Bigr] } \nonumber \\
& & \hspace{1.5cm} + \frac1{6\pi} \Bigl[g_{\mu\nu} u^{\gamma} \partial_{\gamma}
\!-\! 2 u_{(\mu} \partial_{\nu)} \!-\! u_{\mu} u_{\nu} u^{\gamma} 
\partial_{\gamma}\Bigr] \Bigl[ \alpha f - \frac1{\alpha} g^{\rho\sigma} 
\partial_{\rho} \phi \partial_{\sigma} \phi f'\Bigr] \; . \qquad \label{newGmn}
\end{eqnarray}
The final difference is that the auxiliary scalar $\psi[g]$ picks up an extra 
term from the $\chi$ dependence of $\alpha$ \cite{Deffayet:2014XX},
\begin{equation} 
\psi[g] = \frac1{\square}\Biggl( D_{\mu} \Biggl[ \frac{4 \xi g^{\mu\nu}_{\perp} 
u^{\rho} R_{\rho\nu} \!+\! \frac1{3\pi} g^{\mu\nu}_{\perp} \partial_{\nu} 
[\alpha f \!-\! \frac1{\alpha} g^{\rho\sigma} \partial_{\rho} \phi 
\partial_{\sigma} \phi f']}{\sqrt{-g^{\kappa\lambda} \partial_{\kappa} \chi 
\partial_{\lambda} \chi}} \Biggr] \Biggr) \; . \label{newchi}
\end{equation}

These slight changes in the functional form of the field equations conceal vast
differences in their numerical values when specialized to the cosmological
geometry ({\ref{FRW}). For that case the functional $\alpha[g]$ degenerates to
$H(t)/2\pi$. The auxiliary scalars $\phi$, $\chi$ and $\psi$ are unchanged from 
(\ref{FLRWphixi}-\ref{FLRWchipsi}) but $\xi$ is \cite{Deffayet:2014XX},
\begin{equation}
\xi(t) = 2 \!\! \int_{t_i}^t \!\! dt' \dot{\phi}(t') f'\Bigl( \frac{-4\pi^2 
\dot{\phi}^2(t')}{H^2(t')} \Bigr) \; . \label{newxi}
\end{equation}
The MOND addition to the Friedmann equation becomes \cite{Deffayet:2014XX},
\begin{equation}
\Delta G_{00} = -\frac{H^2}{8\pi^2} f\Bigl(\frac{-4\pi^2 \dot{\phi}^2}{H^2}\Bigr)
- \dot{\phi}^2 f'\Bigl(\frac{-4\pi^2 \dot{\phi}^2}{H^2}\Bigr) + 3 H \dot{\xi} + 
6 H^2 \xi \; . \label{newFried}
\end{equation}
Because the argument of the function $f(Z)$ is now nearly constant it is no
longer clear that the reconstruction problem can be solved.

Relation (\ref{alpha}) is probably good enough to 
study the cosmological implications of dynamical $a_0$, but it suffers from 
the obvious problem of vanishing inside any static system. Of course that would 
make the MOND corrections to gravity vanish in precisely the low acceleration,
gravitationally bound systems for which MOND was originally proposed! So a 
more nonlocal version of (\ref{alpha}) must be devised to make the dynamical 
MOND acceleration inside a gravitationally bound system depend upon the 
cosmological acceleration around it.

\section{Conclusions}\label{concl}

There is no question that MOND phenomenology requires any relativistic, 
metric realization of MOND to be nonlocal \cite{Deffayet:2011sk}. Nonlocal
modifications of gravity have been much studied for a variety of reasons
\cite{nonloc}. Section \ref{vacpol} argues that this particular one might 
arise as a residual effect from the quantum gravitational vacuum polarization 
left over from the epoch of primordial inflation. This in no way changes the 
sorts of nonlocal Lagrangians (\ref{general}) one must consider to 
implement MOND phenomenology, but it does justify some of their properties, 
in particular the appearance of an initial time and the fact that gravity 
is modified on large scales rather than small ones. 

Section \ref{blocks} describes the nonlocal building blocks from
which successful models can be constructed. One result of great 
significance is the realization (\ref{B(r)}) that the Newtonian 
potential can be represented as a nonlocal scalar. This means that the
general class of models (\ref{general}) I developed with Cedric Deffayet 
and Gilles Esposito-Farese are a relativistic extension of the Newtonian 
model originally proposed by Bekenstein and Milgrom \cite{Bekenstein:1984tv}. 
It follows that the implications for nonspherical galaxies and strong fields 
have already been worked out from studies of this model.

Section \ref{implement} gives the general field equations 
(\ref{DGmn}-\ref{chipsi}) and their specialization (\ref{FLRWphixi}-\ref{ijeqn})
to the geometry (\ref{FRW}) of cosmology. That completes what might be
termed the ``0th order program'' in developing a relativistic, metric-based
extension of MOND. The ``1st order program'' of subsequent research is
clear:
\begin{itemize}
\item{Apply same techniques that were used for nonlocal cosmology 
\cite{Deser:2013uya,Woodard:2014iga} to check that the nonlocal addition
(\ref{general}) introduces no extra degrees of freedom in addition to those
present in general relativity and that none of the old degrees of freedom 
becomes a ghost; and} 
\item{Solve the reconstruction problem to determine the function $f(Z)$ for
$Z < 0$ needed to support the $\Lambda$CDM expansion history without dark 
matter, both for constant $a_0$ and for dynamical extensions.}
\end{itemize}
If these investigations have a successful outcome one can envisage a 
``2nd order program'' of further studies:
\begin{itemize}
\item{Apply the same techniques that Dodelson and Park developed for
nonlocal cosmology \cite{Park:2012cp,Dodelson:2013sma} to analyze cosmological
perturbations to see if the extra MOND force can make up for the absence of
dark matter in the cosmic microwave background and during structure 
formation;}
\item{Develop a more nonlocal dynamical extension of the MOND acceleration 
which reduces to $a_0$ inside gravitationally bound systems; and}
\item{See if dynamical extensions of $a_0$
can resolve the problems MOND has in reproducing the dynamics of galactic
cluster cores \cite{Aguirre:2001fj} and in explaining recently disturbed 
systems such as the Bullet Cluster \cite{Clowe:2006eq}.}
\end{itemize}

\centerline{\bf Acknowledgements}

I have benefited from years of discussion on this 
subject with C. Deffayet, S. Deser, G. Esposito-Farese, P. Ferreira, S. 
McGaugh, M. Milgrom, R. H. Sanders, C. Skordis, M. A. Soussa and N. C. 
Tsamis. This work was partially supported by NSF grant PHY-1205591, by
the "ARISTEIA" Action of the  "Operational Programme Education and 
Lifelong Learning" and is co-funded by the European Social Fund (ESF) 
and National Resources, and by the Institute for Fundamental Theory at 
the University of Florida.

\end{document}